\IfFileExists{ijcai26.sty}{%
  \documentclass{article}
  \pdfpagewidth=8.5in
  \pdfpageheight=11in
  \usepackage{ijcai26}
}{%
  \documentclass[twocolumn]{article}
  \usepackage[letterpaper,margin=0.75in,columnsep=0.25in]{geometry}
  \makeatletter
  \providecommand{\affiliations}[1]{\par\normalsize ##1}
  \providecommand{\emails}[1]{\par\texttt{\small ##1}}
  \makeatother
}
\usepackage[hidelinks]{hyperref} 

\usepackage{tabularx}
\usepackage{times}
\usepackage{amsmath,amssymb}
\usepackage{booktabs}
\usepackage{graphicx}
\usepackage{tikz}
\usepackage{tikz}
\usetikzlibrary{arrows.meta,positioning,fit}

\usetikzlibrary{arrows.meta,positioning}
\usepackage[ruled,vlined,linesnumbered]{algorithm2e}
\usepackage{url}

\newcommand{\pass}[1]{\ensuremath{\mathrm{Pass}^{#1}}}
\newcommand{\act}{\ensuremath{a}}
\newcommand{\hist}{\ensuremath{h}}

\title{From Natural Language Policies to Executable Obligations: A Verification Harness for Dependable In-Car LLM Agents}

\author{
Radouane Bouchekir, Damir Safin and Tomas Bueno Momcilovic \\
\affiliations
fortiss GmbH, Munich, Germany \\
\emails
\{bouchekir, safin, momcilovic\}@fortiss.org
}
\date{}

\begin{document}
\maketitle

\begin{abstract}
Large Language Models (LLMs) agents deployed in vehicles must satisfy a written operating policy on every turn: a single hallucinated identifier, omitted mandatory side-effect, or premature completion claim fails the task. We present \texttt{AgentGuardUtil}, our entry to CAR-bench Track~1, which treats the AI planer (LLM) as a fallible proposer inside a grounded verify-and-revise loop. Its core novelty is a \emph{runtime policy compiler}: the natural-language policy shipped with each conversation is compiled, once per policy, into typed machine-checkable rules, a subset of which receive an \emph{executable} form. A deterministic obligation engine interprets these rules against live tool results and the simulated
post-write state of the draft itself, emitting the exact remedial calls with computed arguments rather than natural-language reminders. Around this engine, 25 deterministic gates (\textit{identifier provenance, schema and enum validity, gather-before-act, confirmation and future-time protocols}) and an LLM critic produce tiered findings that drive a bounded revision loop tuned for the $\pass{k}$ metric. 
\end{abstract}

\section{Introduction}
LLM-based agents are moving rapidly from research prototypes into products that act on the physical world. In-vehicle voice assistants are an exemplary setting: the agent controls windows, climate, lighting, navigation, charging, and communication on behalf of a driver whose attention is limited, under a manufacturer policy that encodes safety and interaction rules. Therefore, the most impactful scenario involving an unreliable agent is an autonomous action (e.g., a wrong interpretation of a speed limits) that leads to a safety incident (e.g., crash into another vehicle) - far more than a lower benchmark point.

The conditions under which current AI agents are evaluated differ substantially from those encountered in deployment. In operational settings, user requests are frequently incomplete or ambiguous, requiring agents to either infer missing information or request clarification. Necessary tools, parameters, or external resources may be unavailable, making it essential for agents to acknowledge these limitations rather than fabricate outcomes. In addition, deployment environments impose domain-specific policies that regulate how tasks may be performed, including preconditions, mandatory side effects, confirmation requirements, and disclosure obligations. CAR-bench \cite{kirmayr-etal-2026-car} captures these deployment challenges in the in-car domain. The benchmark evaluates ambiguity resolution (\emph{disambiguation}), admitting unavailable capabilities (\emph{hallucination}), and policy-compliant execution (\emph{base}), using conjunctive success criteria over environment state, tool execution, policy compliance, and task termination.

\begin{figure*}[!t]
    \centering
    \includegraphics[
        width=\linewidth,
        height=1\textheight,
        keepaspectratio
    ]{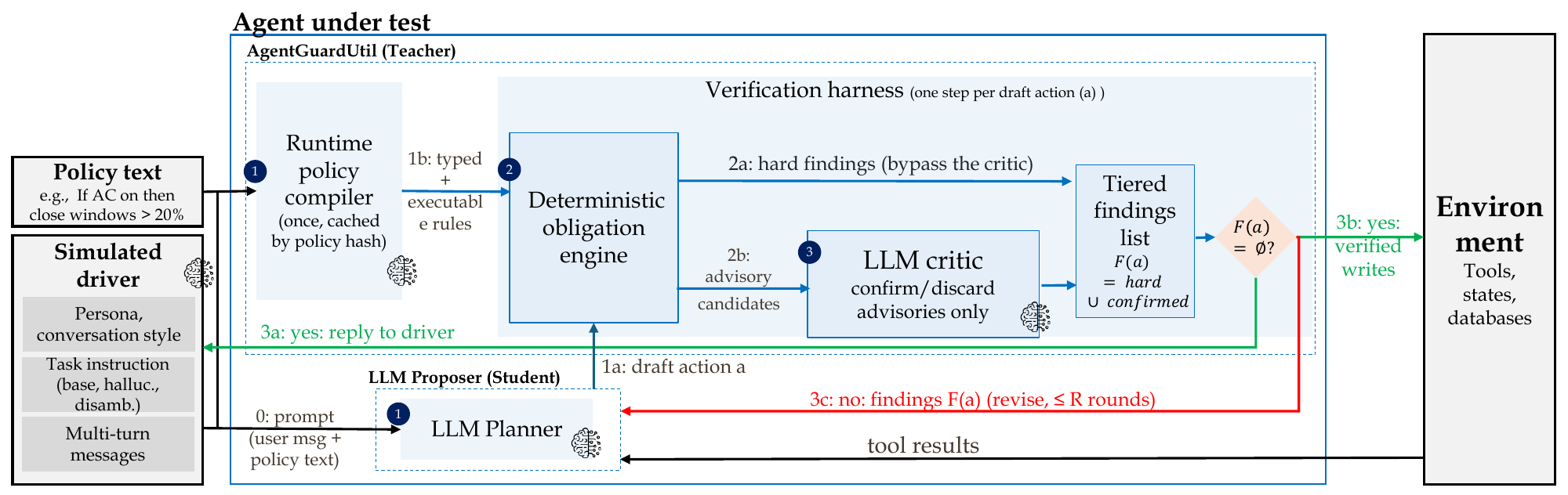}
    \caption{Overview of AgentGuardUtil architecture and one verified turn. A student LLM proposes; a deterministic teacher disposes. The policy is compiled once into executable rules (1). Each draft action (2) passes one verification step: deterministic gates, including the obligation engine, emit hard findings, and an LLM critic may only confirm or discard advisory candidates (3). If the tiered list $F(a)$ is non-empty, the draft returns to the student for revision, $\le$ R rounds; if empty, verified writes reach the environment and the reply reaches the simulated drive. Under the assume–guarantee reading, the student is assumed only capable-but-fallible; every arrow leaving the teacher carries the guarantee.}
    \label{fig:ag_arch}
\end{figure*}

In this paper we present \texttt{AgentGuardUtil}, our submission to CAR-bench Track 1\footnote{\url{https://car-bench.github.io/car-bench/index.html}}. The design rests on two complementary principles (see Fig.~\ref{fig:ag_arch}). First, a \emph{student--teacher} decomposition separates proposal from approval: the LLM (student) only \emph{proposes} actions, while a verification harness (teacher) reviews each draft, returns concrete findings, and re-verifies the revision before any action reaches  the agent (simulated driver). Second, the interaction between the two follows an \emph{assume--guarantee}
discipline in the spirit of compositional verification~\cite{bouchekir2018learning}: the verification harness assumes only that the student is a capable but fallible proposer, and carries the \emph{guarantee} (i.e. policy compliance, grounding in observed data, and policy adherence) through deterministic checks compiled from the conversation's own artifacts (policy text,
tool schemas, observations), independently of the student's behavior. At the first turn, the natural-language policy is compiled into typed machine-checkable rules, part of which receive an \emph{executable} form. On every draft action, a deterministic obligation engine evaluates these rules against live tool results and the simulated
post-write state of the draft itself, computing the exact remedial calls the policy requires; $25$ further deterministic gates enforce identifier provenance, schema validity, and interaction protocols, and an LLM critic adjudicates the uncertain findings. Surviving findings drive a bounded revise-and-recheck loop before anything reaches the simulated driver. The main contributions of this work are summarized as follows: (i) \textbf{\textit{Runtime policy compiler}} producing typed rules with an executable sub-language, cached per policy hash (\S\ref{sec:compile}); (ii) \textbf{\textit{Deterministic obligation engine}} that evaluates executable rules on a projection of the environment state, including \emph{simulation of the draft's own writes}, and emits exact remedial tool calls
(\S\ref{sec:oblig}); (iii) \textbf{\textit{Tiered verification loop}} combining deterministic gates with an LLM critic  (\S\ref{sec:verify}). The design composes chain-of-verification prompting~\cite{dhuliawala2024chain} and self-consistency~\cite{ahmed2023better}, but moves the decisive checks out of the model and into code that the model cannot hallucinate around;

\section{The AgentGuardUtil}

\subsection{Architecture}
Fig. ~\ref{fig:ag_arch} and Algo. ~\ref{alg:turn} follow one turn through the
verification harness. The turn is prepared in three steps. \emph{First}, at the
first turn of a conversation, \textsc{RuntimePolicyCompiler} translates the policy
text $\Pi$ into typed, executable rules $\mathcal{R}$ (\S\ref{sec:compile}); the
result is cached under $\mathrm{hash}(\Pi)$, so later turns skip this step.
\emph{Second}, \textsc{ExtractAsks} scans the new user message and appends each
explicit request to a ledger $L$ of outstanding asks, against which the completion
gate (\S\ref{sec:verify}) later checks any claim of success. \emph{Third}, the
student proposes a draft $\act$, i.e.\ a candidate answer for the turn, a spoken
reply, tool calls, or both, which is neither shown to the driver nor executed
until it passes verification. If the draft would mutate environment state and
voting is enabled, the same prompt is sampled $n$ times from the same LLM and
\textsc{ModalVote} keeps the \emph{action set} proposed most often (the multiset of
tool calls with normalized arguments), a majority vote in the sense of
self-consistency~\cite{ahmed2023better}.

The draft then enters a verify-and-revise loop of at most $R$ rounds, and each
round separates two kinds of evidence. \emph{First}, the deterministic gates of
\S\ref{sec:verify}, among them the obligation engine of \S\ref{sec:oblig},
produce \emph{hard} findings $F_{\mathrm{hard}}$, which stand as issued.
\emph{Second}, the same gates produce \emph{advisory} candidates
$F_{\mathrm{adv}}$, which \textsc{Critic} (an LLM) may confirm or discard, but can
neither extend with a finding of its own nor use to override a hard one. If the
merged tiered list $F$ is empty, the loop exits; otherwise \textsc{Revise}
re-prompts the student with $F$ injected as an internal note, and the revision
re-enters verification.

Two steps close the turn. \emph{First}, because a revision can itself introduce a
new defect, a safety net follows the loop: if the draft was revised and still
carries hard findings, $\textsc{Revise}_{\mathrm{esc}}$ runs once more on those
deterministic findings alone, optionally under a stronger \emph{escalation} model.
\emph{Second}, \textsc{Sanitize} strips the reply of markup that cannot be spoken,
and the turn is released: verified tool calls reach the environment and the
sanitized reply reaches the simulated driver.

\begin{algorithm}[t]
\caption{\textsc{VerifiedTurn}$(\hist,\mathcal{S})$}
\label{alg:turn}
\scriptsize
\If{first turn for this policy $\Pi$}{
  $\mathcal{R}\leftarrow\textsc{RuntimePolicyCompiler}(\Pi,\mathcal{S})$ \tcp*{cached by $\mathrm{hash}(\Pi)$}}
$L \leftarrow L \cup \textsc{ExtractAsks}(\text{new user msg})$\;
$\act\leftarrow\textsc{ModalVote}\big(\{\textsc{LLM}(\hist,\mathcal{S})\}_{1..n}\big)$\;
\For{$i\leftarrow 1$ \KwTo $R$}{
  $F \leftarrow F_{\mathrm{hard}}(\act,\hist,\mathcal{R})
        \,\cup\, \textsc{Critic}\big(F_{\mathrm{adv}}(\act,\hist,\mathcal{R})\big)$\;
  \lIf{$F=\emptyset$}{\textbf{break}}
  $\act\leftarrow\textsc{Revise}(\hist,\act,F)$\;
}
\If{$\act$ was revised \textbf{and} $F_{\mathrm{hard}}(\act)\neq\emptyset$}{
  $\act\leftarrow\textsc{Revise}_{\mathrm{esc}}(\hist,\act,F_{\mathrm{hard}}(\act))$
  \tcp*{safety net}}
\Return{$\textsc{Sanitize}(\act)$}
\end{algorithm}

\subsection{Runtime Policy Compilation}
\label{sec:compile}
At the first turn of a conversation, the policy text $\Pi$ is compiled by an LLM into a rule set $\mathcal{R}$. Each rule $r=\langle \mathrm{id},\,\kappa,\,\mathcal{T}_r,\,\rho\rangle$ carries a type
$\kappa$ (\emph{precondition, auto-action, confirmation, prohibition,
disclosure, constraint, selection}), trigger tools $\mathcal{T}_r$ grounded
against the live tool inventory, and an imperative requirement $\rho$.
Compound clauses are split into atomic rules (one condition, one remedy). A
second, focused pass translates state-conditional rules into an
\emph{executable} form
\begin{equation}
e_r=\big\langle g,\;\varphi,\;\lhd,\;v,\;\omega \big\rangle,
\label{eq:exec}
\end{equation}
where $g$ is a read tool, $\varphi$ a field-name pattern with one wildcard, $\lhd\in\{<,\le,>,\ge,=,\ne,\ni,\not\ni\}$, $v$ a threshold, and $\omega(x)=\langle u,\beta(x)\rangle$ an obligation template binding the matched item $x$ into the arguments of tool $u$. The compiler sees real signatures $u(p_1,\dots,p_m)$; an obligation whose arguments cannot be validated is dropped, degrading that rule to advisory, the engine can only \emph{add} precision, never force an invalid call.

\paragraph{Example.}
\textit{ To illustrate policy compilation, we consider the example of the policy \textbf{``When setting the air conditioning to ON, automatically close all windows if they are open more than 20\%''}.  first compiles to the typed rule $r_{011}=\langle 011,\ \textit{auto-action}, \{\texttt{set\_air\_conditioning}\},\ \rho\rangle$. The trigger set contains
the tool whose invocation \emph{activates} the rule, which is the remedy. The second pass grounds $r_{011}$ into the executable form of Eq.~\eqref{eq:exec}:
\begin{align*}
e_{011}=\big\langle\;&\texttt{get\_vehicle\_window\_positions},\\[-2pt]
&\texttt{window\_*\_position},\ >,\ 20,\\[-2pt]
&\texttt{open\_close\_window}(\langle\mathrm{item}\rangle,\,0)\;\big\rangle.
\end{align*}
The executable form means: fetch the window positions with $g$; every returned field matching $\varphi$ whose value exceeds $v{=}20$ binds its wildcard match (\texttt{driver}, \texttt{passenger}, \dots) into the obligation template $\omega$, validated against the declared enum of the \texttt{window} parameter (\texttt{DRIVER}, \texttt{PASSENGER}, \dots,
\texttt{ALL}). At runtime (\S\ref{sec:oblig}), with observed positions $25/100/0/10$, exactly two obligations survive, $\texttt{open\_close\_window}(\texttt{DRIVER},0)$ and
$\texttt{open\_close\_window}(\texttt{PASSENGER},0)$, and are quoted to the student verbatim with their evidence, \texttt{window\_driver\_position}$\,{=}\,25>20$. }

\subsection{Executable Obligations}
\label{sec:oblig}
The agent perceives the environment only through tool calls: \emph{observation} tools report part of the environment state, and \emph{action} tools change it. For every draft run, the engine answers one question: \emph{given everything the agent has observed and the actions it now proposes, which further actions does the policy still require?} It proceeds in three steps:

- \emph{Step 1: estimate the post-action environment state.} Let $\sigma_g$
be the most recent observation returned by tool $g$, let $w_1,\dots,w_m$
be the actions executed after that observation, and let $W_\act$ be the
actions the draft proposes. Each action may have changed, or would
change, what was observed, so the engine applies all of them, in order,
to the observation and obtains a \emph{projected} state
$\hat{\sigma}=\pi(\sigma_g;\,w_{1..m}\!\cdot\!W_\act)$: its best estimate
of the environment \emph{after} the draft would execute. An action whose
effect is predictable from a compiled rule is \emph{simulated}, i.e.\ the
affected state variable is set to the action's argument value (an
\texttt{ALL}-style argument affects every matching variable); an action
whose effect is not predictable, or that failed, marks the affected
variables as unknown. Because rules are evaluated on this post-action
estimate, a draft whose own actions would create a violation immediately
produces the actions that repair it.

- \emph{Step 2: compute the missing actions.}

A rule is \emph{active} if one of its trigger tools was already
used or appears in the draft
($\mathcal{R}_{\triangleright}=\{r:\mathcal{T}_r\cap(W_\hist\cup
W_\act)\neq\emptyset\}$). 

\begin{equation}
\Omega(\act,\hist)=
\bigcup_{r\in\mathcal{R}_{\triangleright}}
\big\{\omega(x)\;:\;\varphi\!\sim\!x,\ \hat{\sigma}(x)\lhd v\big\}
\;\setminus\;\big(E_\hist\cup A_\act\big).
\label{eq:oblig}
\end{equation}

For every active rule, and for every entity $x$
(a window, a seat, a destination) whose projected value violates the
rule's condition, the engine fills the rule's action template $\omega$
with $x$; it then removes actions the agent has already performed
($E_\hist$) or already included in the draft ($A_\act$). What remains,
$\Omega$, is exactly the set of actions the agent still owes.

- \emph{Step 3: match the expected granularity.} If \emph{every} entity
violates the condition and the tool accepts an \texttt{ALL}-style
argument, the per-entity actions collapse into one collective action, 
the granularity against which the benchmark scores intermediate states.
The mirror check flags \emph{excess}: actions on entities the condition
exempts, and collective actions where only some entities qualify. Each
finding quotes its actions ready to execute, together with the observed
evidence, so the student no longer has to \emph{recall} the policy --- it
only has to \emph{transcribe} the correction.

\paragraph{Example.}
\textit{
Continuing $e_{011}$: the windows were observed at $25/100/0/10$ (driver,
passenger, and the two rear windows), and the draft proposes only
\texttt{set\_air\_conditioning(on)}. That proposed action activates rule
011; no other action has run since the observation, so the projected state
equals the observed one. The driver ($25$) and passenger ($100$) windows
violate the ${>}\,20$ condition, and no closing action exists in the
history or the draft, so
$\Omega=\{\texttt{open\_close\_window}(\texttt{DRIVER},0),\allowbreak\
\texttt{open\_close\_window}(\texttt{PASSENGER},0)\}$ --- issued as a hard
finding with the evidence \texttt{window\_driver\_position}${=}25>20$.
The projection matters in both directions. Had the draft instead proposed
opening \emph{all} windows to 50\% with the air conditioning on,
simulating it would put every window at 50 in the projected state and the
finding would demand the closes --- a violation the draft itself would
create. Conversely, once the student adds the two closing actions,
simulation drives both windows to 0, $\Omega=\emptyset$, and the draft is
approved.}
\begin{table*}[!t]
\centering
\footnotesize
\setlength{\tabcolsep}{5pt}
\renewcommand{\arraystretch}{1.25}
\begin{tabularx}{\textwidth}{@{}l >{\raggedright\arraybackslash}X >{\raggedright\arraybackslash}X@{}}
\toprule
\textbf{Gate} & \textbf{What it checks} & \textbf{Example finding} \\
\midrule

\emph{Identifier provenance} 
& Collects every opaque identifier in the draft's tool calls and looks each one up
  among the identifiers returned by earlier observations. An identifier that was
  never observed blocks the offending call: the agent may \emph{copy} identifiers,
  never \emph{invent} them.
& Routing to \texttt{loc\_lux\_222378} is allowed only if some earlier tool result
  returned that id --- a deterministic anti-hallucination check. \\

\emph{Schema validity} 
& Compares every argument of every proposed call against the tool's declared
  parameter types and allowed values, before anything executes. An argument
  outside the declaration is rejected.
& \texttt{open\_close\_window(window=FRONT)} is rejected because \texttt{FRONT} is
  not among the declared window values. \\

\emph{Gather-before-act} 
& For every rule the draft activates, checks whether the observation its condition
  depends on has ever been made. If not, demands that observation --- always a
  read-only call, never an action.
& Rule 011 conditions on window positions, so \texttt{set\_air\_conditioning(on)}
  is held until \texttt{get\_vehicle\_window\_positions} has been called
  (\S\ref{sec:compile}). \\

\emph{Confirmation protocol} 
& For every proposed call the policy marks as requiring consent, searches the
  history for an intent question that the user affirmed. Without one, the call is
  blocked.
& Opening a window beyond 25\% while the air conditioning is on requires a question
  such as ``shall I proceed?'' and a ``yes'' before the action may execute. \\

\emph{Completion} 
& Extracts the actions the draft's reply claims to have performed and verifies that
  each was actually executed or is part of the draft itself. A claim without a
  matching action is rejected.
& ``I have closed all windows'' with no \texttt{open\_close\_window} call is
  returned to the student. \\

\emph{Refusal-read} 
& When the reply tells the user something cannot be done, checks whether the free
  observation that would answer the request was ever made; if not, demands that
  read first.
& Replying ``I cannot check the rear window'' without ever calling
  \texttt{get\_vehicle\_window\_positions} forces that read. \\

\emph{Future-condition} 
& Detects that the request refers to a future moment (e.g.\ an arrival time) and
  checks that the draft answers it from unfiltered data evaluated at that moment,
  rather than filtering by the current moment.
& ``Will the charging station still be open when I arrive at 20:00?'' must be
  answered from the station's opening hours, not from an \texttt{open\_now} filter
  that reflects the present. \\

\bottomrule
\end{tabularx}
\caption{The principal deterministic gates. Each takes the draft $\act$, the
history $\hist$, and the compiled rules $\mathcal{R}$, and returns findings; all
seven above issue \emph{hard} findings, which stand as issued and always trigger
revision.}
\label{tab:gates}
\end{table*}

\begin{algorithm}[t]
\caption{\textsc{VerifyDraft}$(\act,\hist,\mathcal{R})$ --- one
verification step (the line
$F_{\mathrm{hard}}\cup\textsc{Critic}(F_{\mathrm{adv}})$ of
Algorithm~\ref{alg:turn})}
\label{alg:verify}
\scriptsize
$F_{\mathrm{hard}}\leftarrow\emptyset$;\quad
$F_{\mathrm{adv}}\leftarrow\emptyset$\;
\ForEach{gate $\gamma \in \Gamma$}{
  \tcp{$\Gamma$: $25$ deterministic gates, incl.\ the obligation
       engine $\Omega$ of \S\ref{sec:oblig}}
  $(H,A)\leftarrow\gamma(\act,\hist,\mathcal{R})$
  \tcp*{on internal error: $(\emptyset,\emptyset)$ }
  $F_{\mathrm{hard}}\leftarrow F_{\mathrm{hard}}\cup H$;\quad
  $F_{\mathrm{adv}}\leftarrow F_{\mathrm{adv}}\cup A$\;
}
\If{$F_{\mathrm{adv}}\neq\emptyset$}{
  $F_{\mathrm{adv}}\leftarrow
     \textsc{Critic}(F_{\mathrm{adv}},\hist,\mathcal{S})$
  \tcp*{LLM critic: confirm or discard only; never adds, never
        overrides $F_{\mathrm{hard}}$}}
\ForEach{$f\in F_{\mathrm{hard}}\cup F_{\mathrm{adv}}$}{
  \lIf{$f$ re-demands an action set the student already revised away}{
    drop $f$ and relax the gather gate \tcp*[f]{oscillation valve}}}
\Return{$F_{\mathrm{hard}}\cup F_{\mathrm{adv}}$}
\tcp*{tiered findings list $F(\act)$}
\end{algorithm}

\subsection{Tiered Verification and Variance Control}
\label{sec:verify}
Around the obligation engine run $25$ deterministic gates. Each gate checks one
protocol the draft must respect, and each is decidable from the episode's own
artifacts: the transcript, the tool schemas, and prior observations. Each gate is
a function that takes the draft $\act$, the conversation history $\hist$, and the
compiled rules $\mathcal{R}$, and returns findings, hard or advisory; the most
important ones are summarized in Table~\ref{tab:gates}.

Algorithm~\ref{alg:verify} details how the gates are combined into the single
verification step that Algorithm~\ref{alg:turn} invokes once per revision round.
\emph{First}, every gate $\gamma\in\Gamma$ is run over the same triple
$(\act,\hist,\mathcal{R})$ and its two output sets are accumulated separately; a
gate that raises an internal error contributes nothing, so a defective check can
never block a well-formed draft. \emph{Second}, if any advisory candidates were
raised, the LLM critic adjudicates them as a batch against the transcript and the
tool schemas: it may confirm or discard, but it may neither introduce a finding of
its own nor overturn a hard one. \emph{Third}, an \emph{oscillation valve} scans
the surviving findings and drops any that re-demand an action set the student has
already revised away, relaxing the gather gate for the remainder of the turn; this
prevents two gates from trading a draft back and forth until the round budget $R$
is exhausted. What the step returns is the tiered findings list $F(\act)$.

Tiering is what keeps the harness's own variance bounded. A hard finding is
deterministically certain and always triggers revision, so the decisive checks
never depend on a sampled judgment; an advisory candidate is uncertain by
construction and reaches the student only if the critic confirms it against the
episode's artifacts. Because the critic is confined to filtering a set it did not
generate, a bad sample can only cost a missed advisory, never a spurious
correction or a suppressed hard finding, the guarantees carried by the teacher
degrade gracefully rather than inverting.

\section{Evaluation}
\label{sec:eval}

We report the official CAR-bench Track~1 results\footnote{\url{https://car-bench.github.io/car-bench/leaderboard.html?team=track_1__team-16\#explore}} on the organizers' hidden set
of $30$ tasks, three independent trials each, with the student instantiated as
Claude Opus~4.8 and the harness at its defaults. The primary metric $\pass{3}$
credits a task only when all three trials succeed.

\begin{table}[h]
\centering
\footnotesize
\setlength{\tabcolsep}{4pt}
\begin{tabular}{@{}lccc@{}}
\toprule
& $\pass{3}$ & Pass@3 & $\pass{1}$ \\
\midrule
\texttt{AgentGuardUtil} & $53.3$ & $80.0$ & $66.7$ \\
Organizer baseline      & $50.0$ & $66.7$ & $60.0$ \\
\midrule
\emph{hallucination}    & $70.0$ & $100.0$ & $90.0$ \\
\emph{base}             & $50.0$ & $60.0$  & $60.0$ \\
\emph{disambiguation}   & $40.0$ & $80.0$  & $50.0$ \\
\bottomrule
\end{tabular}
\caption{CAR-bench Track~1 results (\%), in total and per category
($10$ tasks each). $\pass{1}$ is the success rate of a single try. Pass@3
counts a task if at least one try succeeds. $\pass{3}$ counts it only if all
three succeed. Our entry is 9.--11. of $21$ on $\pass{3}$.}
\label{tab:results}
\end{table}

We beat the baseline on all three scores. The gain is large on Pass@3, at
$13.3$ points, and small on $\pass{3}$, at $3.3$ points. This tells us what the
harness does and does not do. It makes a good try better, because it removes
mistakes before they reach the driver. It does not make the three tries more
alike, because the student still writes a different draft each time.

The three categories show where our checks work. \emph{Hallucination} is our
best result. Here the checks are simple facts that code can look up. Was this
identifier seen in an earlier tool result? Does this argument match the tool
schema? No judgment is needed, so the answer is always the same.
\emph{Disambiguation} is our worst result. Code can check that the agent asked
the user a question and got a ``yes''. Code cannot tell whether the agent
\emph{should} have asked, because nothing in the transcript or the schemas
answers that. This is left to the LLM critic, and the critic is not always
right.

The main loss is the gap between Pass@3 ($80.0$) and $\pass{3}$ ($53.3$). Eight
tasks succeed in some tries but not in all three. Our checks are stable, but the
student is not, and the checks are silent when no compiled rule covers the task.
So the way to raise $\pass{3}$ is to turn more of the policy into executable
rules. The harness is also slow and expensive. One try costs $362$k tokens and
$19.2$\,s per turn, against $82$k and $3.2$\,s for the baseline. Across the
whole field, spending more tokens does not give a better score
($r{=}0.435$).  

\section{Conclusion}
\texttt{AgentGuardUtil} pairs an LLM student that drafts each turn with a deterministic teacher that checks every draft before it takes effect: the policy is compiled once into executable rules, each draft is evaluated against what was actually observed, including the state the environment would reach after the draft's own actions, and the findings go back to the student until none remain. This addresses the three challenge classes directly: hallucinated identifiers are blocked (every outgoing identifier must have been observed), ambiguous requests are held behind confirmation checks, and forgotten policy side-effects come back as ready-to-execute actions. 


\bibliographystyle{plain}
\bibliography{ref}

@article{bouchekir2018learning,
  title={Learning-based symbolic assume-guarantee reasoning for Markov decision process by using interval Markov process},
  author={Bouchekir, Redouane and Boukala, Mohand Cherif},
  journal={Innovations in Systems and Software Engineering},
  volume={14},
  number={3},
  pages={229--244},
  year={2018},
  publisher={Springer}
}

@inproceedings{kirmayr-etal-2026-car,
    title = "{CAR}-bench: Evaluating the Consistency and Limit-Awareness of {LLM} Agents under Real-World Uncertainty",
    author = "Kirmayr, Johannes  and
      Stappen, Lukas  and
      Andre, Elisabeth",
    editor = "Liakata, Maria  and
      Moreira, Viviane P.  and
      Zhang, Jiajun  and
      Jurgens, David",
    booktitle = "Proceedings of the 64th Annual Meeting of the {A}ssociation for {C}omputational {L}inguistics (Volume 1: Long Papers)",
    month = jul,
    year = "2026",
    address = "San Diego, California, United States",
    publisher = "Association for Computational Linguistics",
    url = "https://aclanthology.org/2026.acl-long.1886/",
    doi = "10.18653/v1/2026.acl-long.1886",
    pages = "40599--40618",
    ISBN = "979-8-89176-390-6"
}

@inproceedings{dhuliawala2024chain,
  title={Chain-of-verification reduces hallucination in large language models},
  author={Dhuliawala, Shehzaad and Komeili, Mojtaba and Xu, Jing and Raileanu, Roberta and Li, Xian and Celikyilmaz, Asli and Weston, Jason},
  booktitle={Findings of the association for computational linguistics: ACL 2024},
  pages={3563--3578},
  year={2024}
}

@inproceedings{ahmed2023better,
  title={Better patching using llm prompting, via self-consistency},
  author={Ahmed, Toufique and Devanbu, Premkumar},
  booktitle={2023 38th IEEE/ACM International Conference on Automated Software Engineering (ASE)},
  pages={1742--1746},
  year={2023},
  organization={IEEE}
}

\end{document}